\documentclass[cits]{PoS}

\title{Transient Radio Neutron Stars}

\ShortTitle{Transient Radio Neutron Stars}

\author{\speaker{Evan Keane}\\ Jodrell Bank Centre for Astrophysics,
        School of Physics \& Astronomy, University of Manchester,
        Manchester M13 9PL, UK.\\ E-mail:
        \email{Evan.Keane@gmail.com}}

\abstract{Here I will review the high time resolution radio sky,
focusing on millisecond scales. This is primarily occupied by neutron
stars, the well-known radio pulsars and the recently identified group
of transient sources known as Rotating RAdio Transients (RRATs). The
RRATs appear to be abundant in the Galaxy, which at first glance may
be difficult to reconcile with the observed supernova rate. However,
as I will discuss, it seems that the RRATs can be explained as pulsars
which are either extreme nullers, highly variable or weak/distant. I
will re-cap some recent results including a re-analysis of the Parkes
Multi-beam Pulsar Survey, which has identified several new sources, as
well as the unusual timing behaviour of RRAT J1819$-$1458. This leads
to an examination of where RRATs fit within the evolution of neutron
stars post-supernova.}

\FullConference{High Time Resolution Astrophysics IV - The Era of
		Extremely Large Telescopes - HTRA-IV,\\ May 5-7,
		2010\\ Agios Nikolaos, Crete, Greece}

\begin{document}

\section{Introduction}
Short-timescale bursts, pulses or flickering at radio frequencies
signal extreme astrophysical environments. A pulse of width $W$ with a
flux density $S$ at an observing frequency $\nu$ which originates from
a source at a distance $D$ has a brightness temperature of
\begin{equation}\label{eq:tb}
  T_{\rm{B}}\geq4.152\times10^{23}\;\rm{K}\left(\frac{SD^2}{\rm{Jy.kpc^2}}\right)\left(\frac{\rm{GHz.ms}}{\nu.W}\right)^2.
\end{equation}
The minimum $T_{\rm{B}}$ in this expression is obtained when the
emitting region is the maximum size of a causally connected region
$cW=300\;\mathrm{km}(W/1\;\mathrm{ms})$. Equation~\ref{eq:tb} is
parameterised in units typical of Galactic millisecond bursts which we
will discuss below. Thus observations of the transient radio sky probe
compact objects and coherent non-thermal emission processes. If the
dynamical time $t_{\mathrm{dyn}}=\sqrt{1/G\rho}$ dictates the scale on
which we see changes then the millisecond radio sky consists mainly of
neutron stars which have $t_{\mathrm{dyn}}\sim0.1\;\mathrm{ms}$.

Neutron stars are the most populous member of `transient phase space'
\cite{clm04} and it is on them that this work focuses. We will discuss
the well known radio pulsars (see e.g.~\cite{ls05a,lk05}) which have
been joined in recent times by the `intermittent
pulsars'~\cite{klo+06} and the `RRATs' (eRRATic radio sources, aka
Rotating RAdio Transients,~\cite{mll+06}). Together these sources
exhibit variability on timescales spanning 16 orders of
magnitude. Giant pulses of nanosecond duration have been observed in
the Crab pulsar~\cite{hkwe03}, whereas PSR~B1931+24 has been seen to
regularly switch on and off for $\sim5$ and $\sim30$ days respectively
($\sim10^7$~s)~\cite{klo+06}. The discovery of RRATs in particular has
sparked much interest in radio transients. Their inferred population
is large and a number of systematic searches of pulsar survey data
have been performed~\cite{mll+06,hrk+07,dcm+09,kle+10,bb10}, all of
which have been successful in identifying new sources. In this paper
we ask the question of whether RRATs are distinct from radio pulsars
and the other manifestations of neutron stars with the aim of deciding
where they fit in the `neutron star zoo'. We begin by discussing in \S
2 the transient radio behaviour of the different neutron star
classes. \S 3 then discusses the Galactic birthrates of neutron stars
and how they compare to the observed core-collapse supernova rate in
the Galaxy, in light of a recent re-processing of the Parkes
Multi-beam Pulsar Survey (PMPS). We then, in \S 4, discuss the current
state of knowledge regarding the evolution of neutron stars
post-supernova, before concluding in \S 5.

\section{Transient Neutron Stars}

\subsection{Basic Pulsar Model}
The standard model of a pulsar is a rapidly spinning neutron star with
a dipolar magnetic field emitting a coherent beam of radio emission
along its magnetic poles powered by the loss of rotational energy
\cite{lk05}. If, as it rotates, the pulsar beam cuts our line of
sight, a highly periodic source is detectable at the Earth. A typical
pulsar has a period of $P\sim0.5$~s and slows down at a rate of
$\dot{P}\sim 10^{-15}$. Such sources are referred to as `normal' or
`slow' pulsars and comprise the majority of the $\sim1800$ presently
known radio pulsars. In addition to these there are the millisecond
pulsars (MSPs) which are the fastest rotators with typical periods of
a few milliseconds. 

\begin{figure}[h]  
  \begin{center}
    \includegraphics[scale=0.6,angle=-90]{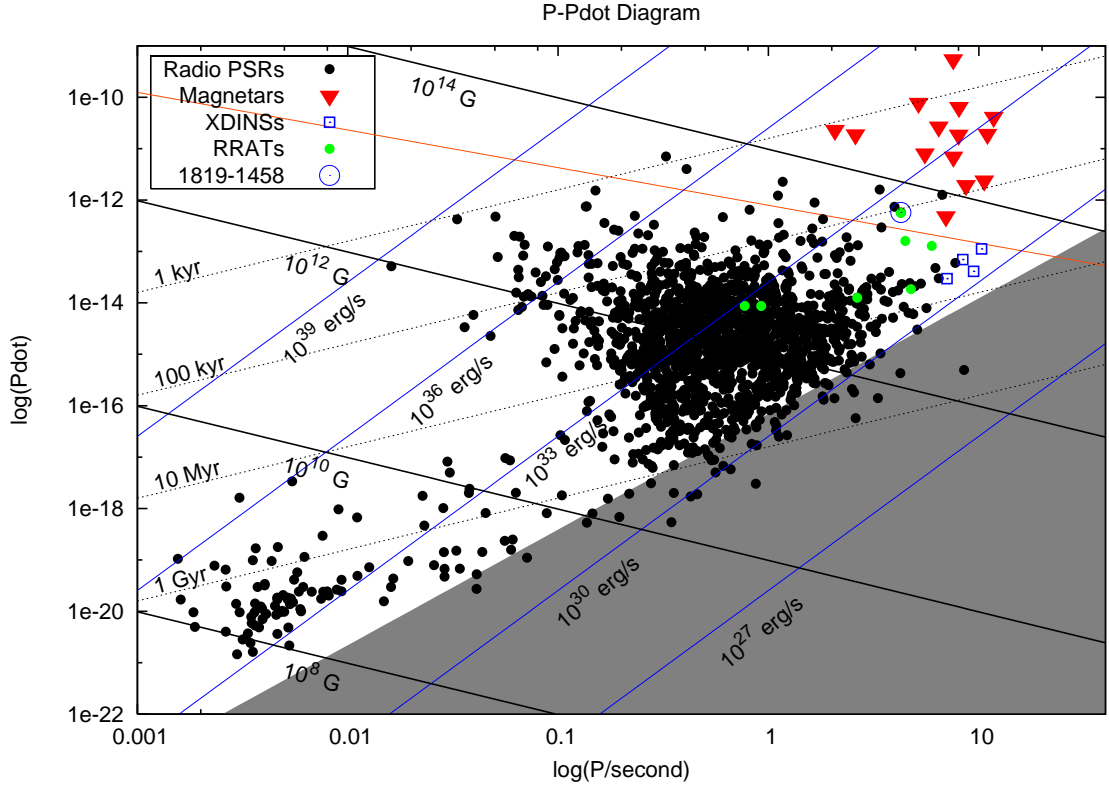} 
  \end{center}      
  \caption{\small{The pulsar $P-\dot{P}$ diagram. Shown are the radio
  pulsars, which can clearly be seen to consist of two classes --- the
  `slow' pulsars and the MSPs, as well as those RRATs (J1819$-$1458 is
  circled), XDINSs and magnetars with known period derivative. The
  shaded region in the bottom right denotes the canonical `death
  valley' of \cite{cr93a} where we can see there is a distinct lack of
  sources. The radio loud-radio quiet boundary of \cite{bh98} is also
  shown and we can see that only $\sim1\%$ of sources are found above
  this line. Also plotted are lines of constant $B$, $\dot{E}$ and
  $\tau_{\mathrm{c}}$.}}
  \vspace{-15pt}
  \label{fig:ppdot}
\end{figure}

We can derive some simple equations to quantify some pulsar
parameters. The rate of rotational energy loss of a pulsar is simply
$\dot{E}=d/dt[(1/2)I\Omega^2]$. Using canonical neutron star values
for mass ($1.4\;\mathrm{M_{\bigodot}}$) and radius ($10$~km) we can
take the moment of inertia to be that of a sphere to yield
\begin{equation}
	\dot{E}=3.95\times10^{31} \; \mbox{\rm ergs.s}^{-1} \;
	\left( \frac{\dot{P}}{10^{-15}} \right) \;
	\left( \frac{P}{\mbox{\rm s}} \right)^{-3}.
\end{equation}
Equating this energy loss rate to the well known expression for the
loss rate of a rotating magnetic dipole we can obtain an estimate for
the `characteristic magnetic field strength' which is
\begin{equation}\label{eq:B}
  B=1.0\times10^{12}\; \mathrm{G} \;
 \sqrt{ \left(\frac{\dot{P}}{10^{-15}} \right)
 \left( \frac{P}{\mathrm{s}} \right)}.
\end{equation}
Assuming a spin-down law of the form $\dot{P}=KP^{2-n}$ we can
determine an evolutionary timescale for pulsars by considering the
case of a pulsar born spinning at a much faster rate than presently
observed, i.e. $P_{{\rm birth}}\ll P_{{\rm now}}$. The `spin-down'
timescale is given by integrating the spin-down law to get
\begin{equation}
  \tau=\frac{1}{(n-1)}\frac{P}{\dot{P}} \;\;\; \textrm{.}
\end{equation}
For the dipolar case $n=3$ which gives us the `characteristic'
timescale, $\tau_{\mathrm{c}}=P/2\dot{P}$. This is commonly referred
to as the pulsar's `age' however we emphasise that $\tau_{\mathrm{c}}$
is only a true representation of the pulsar age when the above
assumptions are valid. Nevertheless $\tau_{\mathrm{c}}$ does provide
us with a representative timescale for pulsar evolution. Pulsars are
typically classified using a $P-\dot{P}$ diagram as shown in
Figure~\ref{fig:ppdot} and, using the above equations, lines of
constant $\dot{E}$, $B$ and $\tau_{\mathrm{c}}$ are shown on this
diagram.

\subsection{Pulsar Stability}
\begin{figure}  
  \begin{center}
    \includegraphics[trim = 0mm 0mm 15mm 0mm, clip, scale=0.5, angle=-90]{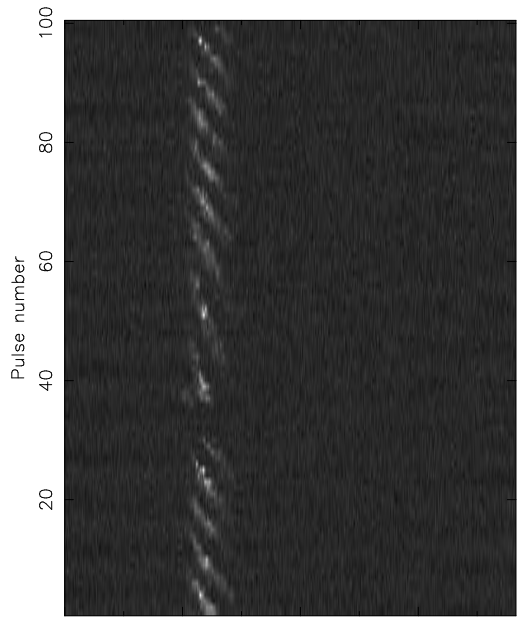} 
  \end{center}   
  \vspace{-15pt}
  \caption{\small{Plotted are a sequence of 100 pulses from
  PSR~B0031$-$07 observed with the Westerboork Synthesis Radio
  Telescope. Periodic drifting of the pulse in pulse longitude is
  evident. This pulsar also exhibits nulls such as that visible
  between the fifth and sixth drift bands. (Image credit:
  M. Serylak).}}
%  \vspace{-15pt}
  \label{fig:drift}
\end{figure}

Pulsars are commonly referred to as stable astrophysical clocks but
this is true only when considering integrated pulse profiles (of
$10^4$ periods or more), especially those of the MSPs. Conversely, on
a period-by-period basis the pulses we detect from pulsars are quite
variable and exhibit much random as well as highly organised
behaviour. Sub-pulse drifting is a phenomenon whereby the rotational
phase where we see pulsar emission changes periodically (see
Figure~\ref{fig:drift}). Some pulsars also exhibit `mode-changing'
whereby they switch between two or more different stable emission
profiles. Nulling can be seen as an extreme example of moding where
one of the modes shows no radio emission, i.e. the radio emission
ceases and the pulsar is `off'. Typical nulling occurs for $1-10$
rotation periods but we must note that the observed selection of
nulling pulsars is quite biased \cite{wmj07}. Pulsars with longer
nulling fraction are less likely to be detected in a single survey
pointing, and in a confirmation pointing, and hence may be discarded
amongst the plethora of pulsar candidates produced in modern
surveys\footnote{The number of pulsar candidates produced in modern
surveys has surpassed what can be inspected by humans in a reasonable
time. This has led to the use of artificial neural networks to
identify the best candidates \cite{emk+10}.}. Also, due to a lack of
sufficient signal-to-noise ratio, weaker pulsars cannot be examined on
shorter timescales. Thus there may well be nulling occurring either
unnoticed or undetectable in many known pulsars.

%To perform `precision pulsar timing'\footnote{Well-spaced reliable
%clocks spread throughout the Galaxy are very useful, e.g. for probing
%strong gravitational fields, see \cite{fhb+10} for a review of recent
%efforts at gravitational wave detection using pulsars.} integrated
%profiles are used which are assumed to be shifted, scaled and noisier
%versions of a template profile intrinsic to the pulsar. A pulse
%time-of-arrival (TOA) is obtained by cross-correlating the observed
%profile with the template. For reliable TOAs the short-term 'sub-pulse
%modulation' effects must be averaged over to provide stable profiles,
%i.e. where the cross correlation coefficent increases with $\sqrt{N}$
%where $N$ is the number of pulses averaged. For MSPs, $\sim10^4$
%pulses must be added to obtain a stable profile which is easily
%achieved. For slower pulsars this stability criterion is not usually
%reached \cite{hmt75,rr95}, i.e. the most precise pulsar timing is
%performed using MSPs.

\subsection{RRATs}
In 2006 eleven new sources, dubbed RRATs, were discovered in an
archival search \cite{mll+06} of the PMPS \cite{mlc+01}. These sources
are characterised by detectable millisecond bursts of radio emission
occurring as infrequently as every 3 hours to as often as every few
minutes. The bursts have a duration of $\sim1-30$~ms with peak flux
densities (at $1.4$~GHz) of $\sim0.1-10$~Jy. In total this amounts to
a mere 5~minutes of detectable radio emission per year for a typical
RRAT, which illustrates the inherent difficulty in detecting such
sources. As the RRATs are located at distances of a few kpc,
Equation~\ref{eq:tb} tells us that the brightness temperatures are
high at $10^{22}-10^{23}$~K, which is within the range measured for
radio pulsars. Observing a number of pulses from each source has
enabled the determination of underlying periodicities using time
differencing methods (see e.g. \cite{kle+10}) with periodicities in
the range $0.7-7$~s for the original 11 sources. The most well studied
source, J1819$-$1458, has been observed in the X-ray where a thermal
spectrum at $kT\sim140$~eV is seen \cite{rbg+06b,mrg+07b,rmg+09}. All
these characteristics point towards RRATs being neutron
stars. Monitoring these sources reveals that their spin periods are
slowing down, and we can measure period derivatives and place them in
$P-\dot{P}$ space (Figure~\ref{fig:ppdot}). Here we can see that they
seem to occupy a region similar to the so-called high-B radio pulsars,
i.e. between the main pulsar population and the magnetars. The arrival
times of the pulses themselves seem to be random, not yet showing any
highly significant quasi-periodicities on timescales up to 1000~days,
although this is limited by the number of detected pulses
\cite{nipuni_inprep}. Given the difficulty in detecting RRATs, we can
estimate the selection effects and make a prediction of the Galactic
population of RRATs. This yields $N_{\mathrm{RRAT}}=\gamma
N_{\mathrm{PSR}}$ with $\gamma=1-3$ \cite{mll+06,kle+10} but contains
at least three built-in sources of uncertainty, namely the burst rate
distribution, the fraction of RRATs obscured from detection due to the
effects of radio frequency interference and the beaming fraction of
RRATs. Of course, this estimate is also extrapolated from a very small
sample population and there may be other unknown selection
effects. Nonetheless, below we take this claim at face value and
investigate the implications for the Galactic neutron star
population. We also describe a re-analysis of the PMPS in search of
more RRAT sources, conducted to better understand their phenomenology
as well as improving the population estimate.
%$f_{\mathrm{ON}}$ $f_{\mathrm{RFI}}$ and $f_{\mathrm{beam}}$ and other
%selection effects?

\subsection{Intermittent Pulsars}
2006 also saw the discovery of `intermittent pulsars', sources which
behave as normal radio pulsars for several days before switching off
entirely for days to weeks. This switching occurs in a quasi-periodic
fashion with the archetypal system PSR~B1931+24 turning `on' for
$5-10$ days and `off' for $25-35$ days \cite{klo+06}. These timescales
allow the measurement of separate slow-down rates during the on and
off states, $\dot{\nu}_{\mathrm{on}}$ and
$\dot{\nu}_{\mathrm{off}}$. The difference in these rates is about
$50\%$ and reflects the extra energy loss due to the pulsar wind when
there is radio emission. When off, the star slows down via dipole
braking alone. When on, it has been seen to turn off in the space of a
few seconds. This indicates a massive change in magnetospheric
currents on a very short timescale to a new state which is apparently
stable for $\sim10^6$ periods before switching once more. The
explanation as to why this switching is quasi-periodic is unknown. We
note that the scenario of two slow-down rates should apply to RRATs
also, if they are truly off (see \S 3 for a discussion of this). When
on, a RRAT slows down at a rate $\dot{\nu_{\mathrm{on}}}$. The
slow-down rate of a RRAT is
$\dot{\nu}_{\mathrm{RRAT}}=\dot\nu_{\mathrm{on}}f_{\mathrm{on}}-\dot\nu_{\mathrm{off}}(1-f_{\mathrm{on}})$
where $f_{\mathrm{on}}$ is the fraction of time the RRAT is on given
by $f_{\mathrm{on}}=gW/f_{\mathrm{beam}}T_{\mathrm{obs}}$. The factor
$g$ is the observed RRAT pulses/period, $W$ is the pulse width,
$f_{\mathrm{beam}}$ is the beaming fraction\footnote{The beaming
fraction factor is necessary as we remember that the RRAT is `on' also
when pointed away from us.} (empirically found to be a function of
period \cite{tm98}) and $T_{\mathrm{obs}}$ is the range over which the
observations were performed. A typical RRAT (see e.g. \cite{kle+10}
for typical numbers) has $f_{\mathrm{on}}\ll 1$ so that
$\dot{\nu}_{\mathrm{RRAT}}\approx\dot{\nu}_{\mathrm{off}}$,
i.e. measuring two slow-down rates is not possible for RRATs unlike in
the case of intermittent pulsars where the timescales are more
favourable.

\subsection{Death Valley}
Pulsar emission requires a supply of particles from the stellar
surface which can be accelerated in the pulsar magnetosphere for pair
production, $\gamma\rightarrow e^+ + e^-$, to ultimately lead to
coherent radio emission. The strength of the electric potential
$\Delta V$ depends on $B$ and $P$, e.g. in the simple Goldreich-Julian
case $\Delta V\propto B/P^2$ \cite{gj69}. An electron accelerated in
this potential will acquire a Lorentz factor of $e\Delta
V/m_{\mathrm{e}}c^2$. Depending on the emission mechanism (i.e. the
dependence of $\Delta V$ on $B$ and $P$) the minimum Lorentz factor
sufficient for pair-production (the photon must have energy of at
least $2m_{\mathrm{e}}c^2$) defines a `death-line', separating regions
of $P-B$ space where radio pulsar emission is possible and regions
where it is inhibited (the `death valley'). Detailed considerations
lead to different death-lines for different field configurations,
e.g. on high curvature field lines \cite{cr93a}, and death-lines for
several emission mechanisms have been proposed (see
e.g. \cite{aro96,qz96,zhm00}). We note however that the various
death-lines do not satisfactorily explain the observed pulsar
population and there is at least one pulsar which flouts the rules in
the death valley, namely the $8.5$-second PSR~J2144$-$3933 whose
detection as a radio pulsar poses serious challenges to pulsar
emission theories \cite{ymj99}.

\section{RRATs: Recent Results}

\subsection{Too Many Neutron Stars?}
As we have mentioned above, the projected Galactic population of RRATs
is large, perhaps larger than the population of radio pulsars. With
this in mind, and considering all the classes of neutron stars now
known, it makes sense to revisit the question as to whether the
numbers are consistent with the observed core-collapse supernova rate
in the Galaxy. From measurements of Galactic $\gamma$-ray emission
from radioactive aluminium this has been determined to be
$\beta_{\mathrm{CCSN}}=1.9\pm1.1\;\mathrm{century}^{-1}$
\cite{diehl+06}, and of course the birthrate of neutron stars cannot
exceed this. If we make the assumption that all the known classes of
neutron stars are independent populations then this condition becomes
\begin{equation}\label{eq:balance}
  \beta_{\mathrm{CCSN}}\ge\beta_{\mathrm{PSR}}+\beta_{\mathrm{XDINS}}+\beta_{\mathrm{RRAT}}+\beta_{\mathrm{magnetar}}+\beta_{\mathrm{CCO}},
\end{equation}
where each rate $\beta_{\mathrm{X}}$ is the birthrate (per century)
for neutron stars of type X. Equation~\ref{eq:balance} refers to
different neutron star populations which we now quickly summarise.

\begin{figure}  
  \begin{center}
    \includegraphics[scale=0.4,angle=-90]{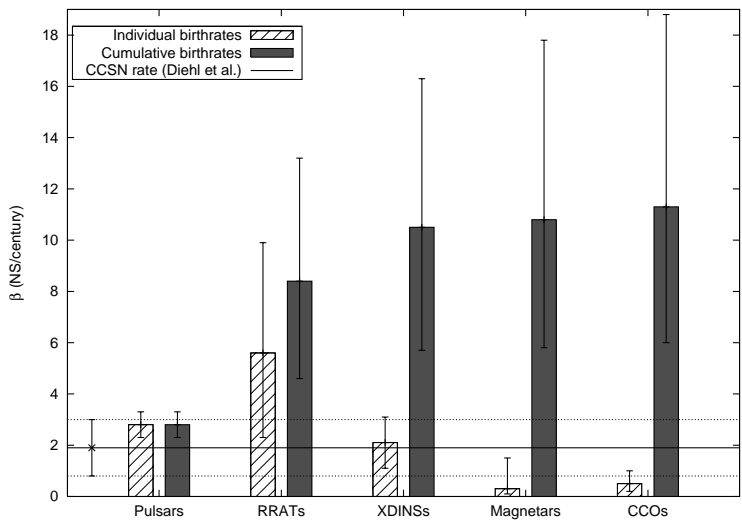} 
  \end{center}   
  \vspace{-15pt}
  \caption{\small{The estimates for individual neutron star birthrates
  for the different populations (hatched boxes), cumulative birthrate
  (solid boxes) and the core-collapse supernova rate (solid
  line). Adapted from a version in \cite{kk08}.}}
%  \vspace{-15pt}
  \label{fig:birthrates}
\end{figure}

%The fact that the majority of MSPs are in binary systems, and are
%$\gtrsim30$ times more likely to be in binaries than slow
%pulsars\footnote{See for example the ATNF pulsar catalogue,
%\texttt{http://www.atnf.csiro.au/research/pulsar/psrcat/}}, supports
%this model.

% FIT THIS IN SOMEWHERE
%The MSPs are `recycled' pulsars having acquired
%their high spin rates via angular momentum transfer from a binary
%partner. At slower periods there are radio pulsars with periods as
%slow as $8.5$~s. Several other manifestations of neutron star are
%known with periods in this range. Firstly there are the Isolated
%Neutron Stars (INSs, aka XDINSs, see e.g. \cite{kk09} and references
%therein) seen only via their thermal radiation. Secondly there are the
%magnetars \cite{wt04} which, until recently, were observed only at
%high energies (X-ray and $\gamma$-ray), but have shown transient
%pulsar-like radio emission in two cases \cite{crh+06,ksj+07}. There
%are also the Central Compact Objects (CCOs, see e.g. \cite{hg10} and
%references therein) supernova remnant associated non-variable thermal
%X-ray point sources thought to be young isolated neutron stars.
%

The (X-ray-Dim) Isolated Neutron Stars (XDINSs, aka INSs, see
e.g. \cite{kk09}) are a group of nearby neutron stars (sometimes
referred to as ``The Magnificent Seven'') seen only via their thermal
emission. There have been extensive searches for radio emission from
XDINSs with no detection \cite{kml+09} and their X-ray spectra are
well fit as blackbodies, without the need for a power-law component,
which would be suggestive of an active magnetosphere.

The magnetars consist of the Soft Gamma Repeaters (SGRs) and the
Anomalous X-ray Pulsars (AXPs) \cite{wt04}. These are thought to be
isolated neutron stars with very strong magnetic fields of
$10^{14}-10^{15}$~G whose emission is powered by magnetic field decay.
These magnetic fields exceed the `quantum critical field'
strength\footnote{The quantum critical value is that which makes the
energy gap of electron cyclotron orbits (`Landau levels') equal the
electron rest mass. In SI units $\Delta E=\hbar eB/m_{\mathrm{e}}$ so
that $B_{\mathrm{QC}}=m^2c^2/q\hbar$.}
$B_{\mathrm{QC}}=4.4\times10^{13}$~G so that higher order Quantum
Electrodynamics effects play a role. For example, the amplitude for
photon splitting, $\gamma\rightarrow\gamma+\gamma$, a third order
effect, is proportional to
$\alpha^3(\hbar\omega/m_{\mathrm{e}}c^2)^5(B/B_{\mathrm{QC}})^6$
\cite{adl71} where $\alpha$ is the fine structure constant and
$\hbar\omega$ is the photon energy. In magnetic fields $\gtrsim
B_{\mathrm{QC}}$ this dominates over photo-pair creation quenching the
build-up of plasma and hence the radio emission \cite{bh98}. For many
years the known magnetars (radio-quiet) and the pulsars (radio-loud)
were well separated into regions where this process was dominant or
suppressed respectively. However, recently some magnetars have been
found to be radio-loud \cite{cmh+06,crj+08} and several radio pulsars
with $B>B_{\mathrm{QC}}$ have been identified (see
Figure~\ref{fig:ppdot}). So, although there is a dearth of sources in
the $B\sim B_{\mathrm{QC}}$ region, the fact that there are any at all
implies that photon splitting may not always dominate over
pair-creation, e.g. if single polarisation selection rules forbid it
\cite{wm06}.

%Despite these recent results
%magnetars remain primarily studied at high energies.

The Central Compact Objects (CCOs) are another small group of neutron
stars which are isolated point sources associated with supernova
remnants and are seen in thermal X-rays \cite{hg10}. CCOs have no
optical or radio counterparts and do not have associated pulsar wind
nebulae. Recently the first measurement of a period derivative for a
CCO has been performed for PSR~J1852+0040, associated with the SNR
Kestevan 79, which has $\dot{P}=(8.68\pm0.09)\times10^{-18}$
\cite{hg10}, implying, in the dipolar magnetic field scenario, the
lowest magnetic field strength of any young neutron star of just
$B=3.1\times10^{10}$~G. An estimate of their birthrates was made by
\cite{gbs00} to be
$\beta_{\mathrm{CCO}}\approx0.5\;\mathrm{century}^{-1}$. Although
excluded from the initial argument \cite{kk08}, we include the CCO
birthrate in Figure~\ref{fig:birthrates} with the addition of an ad
hoc uncertainty factor of 2.

One point of clarification is that $\beta_{\mathrm{PSR}}$ is the
birthrate of `normal' pulsars only, i.e. not the MSPs. This is
because, according to the standard evolutionary picture \cite{acrs82},
normal pulsars (in binary systems) are progenitors of MSPs. After
$\sim10^7$~yr, once a pulsar has slowed down and crossed the death
line, it will no longer act as a radio pulsar. However, if this `dead'
pulsar has a binary companion it can experience a re-birth. Accretion
from the companion re-heats areas of the neutron star surface and
periodic X-ray emission will be visible from these hot spots. The
system is now a low-mass X-ray binary (LMXB). In addition to mass
transfer there is a transfer of angular momentum and the dead star is
spun up to spin frequencies of hundreds of Hz and re-activated as a
radio pulsar --- an MSP. The transition from LMXB to MSP has been
observed in PSR~J1023+0038 over the last decade \cite{asr+09}.

Using the best estimates for the various birthrates (see \cite{kk08}
and references therein), and assuming that the observed classes are
distinct populations, we conclude that the supernova rate cannot keep
pace with the necessary rate of neutron star production. This seems to
point out that our assumption of distinct populations is invalid and
has led naturally to the suggestion that the various classes are
linked \cite{kk08}. This link may be evolutionary in any of a few
senses: (1) pulsars may evolve so as to increase nulling; (2) once a
pulsar has evolved to particular areas of parameter space the
selection effects may be changed so that the source appears more
sporadic; (3) RRAT emission may be an extra `mode' of emission in
addition to the more steady (i.e. without nulling) emission seen in
slow pulsars. Evolving beams would change the `mode' in which we see
the source during its lifetime. These possibilities are discussed in
more detail in \S 4.

An alternative explanation to the `birthrate problem' might simply be
that the birthrate estimates are incorrect. This could allow the
possibility of distinct populations (see \S 4 for a discussion of
this). While this solution may seem less satisfactory, it can be
directly investigated. The RRAT population estimate is the obvious
target to try to improve: they make a large contribution to the
putative birthrate problem and there is much survey data which has not
been exhaustively searched for RRATs wherein many more may be
discovered. With the discovery of many more RRATs an improved
population estimate (and hopefully much other understanding) would
follow.

\subsection{PMSingle}
With the goal of discovering more RRATs in the PMPS a complete
reprocessing was recently performed, an analysis referred to as
PMSingle \cite{kle+10}. This doubled the known PMPS RRATs to 22 with a
few more confirmations soon to be reported \cite{keane_inprep}. Using
radio frequency interference (RFI) removal techniques as described in
\cite{ekl09} this search effectively set the fraction of sources
missed by RFI to zero, thus removing this source of uncertainty
\cite{mll+06} from the RRAT population estimate. The PMPS used a
13-beam receiver at $1.4$~GHz where it is unlikely that a true
astrophysical source would show up in more than a single beam, unless
extremely bright (e.g. \cite{lbm+07}). In addition to the RFI removal
techniques, this re-analysis rejected multi-beam sources on this basis
as well as expanding some of the parameter space searched (e.g. for
wider pulses). Continued monitoring of the newly identified RRATs will
result in a determination of the RRAT burst rate distribution and
efforts to this end are ongoing \cite{keane_inprep}. With this
information we can perform a detailed population synthesis of RRATs
but for now we can say that the new discoveries are consistent with
the initial population estimate, i.e. there do seem to be about as
many RRATs as radio pulsars in our Galaxy. However this statement must
be interpreted carefully as described in \S 4. This effort to identify
new RRATs is being helped by other searches which have also identified
numerous sources, in surveys at GBT \cite{hrk+07}, Arecibo
\cite{dcm+09} and archival searches of higher latitude Parkes surveys
\cite{bb10}.

\subsection{Unusual Glitches}
\begin{figure}  
  \begin{center}
    \includegraphics[scale=0.4,angle=0]{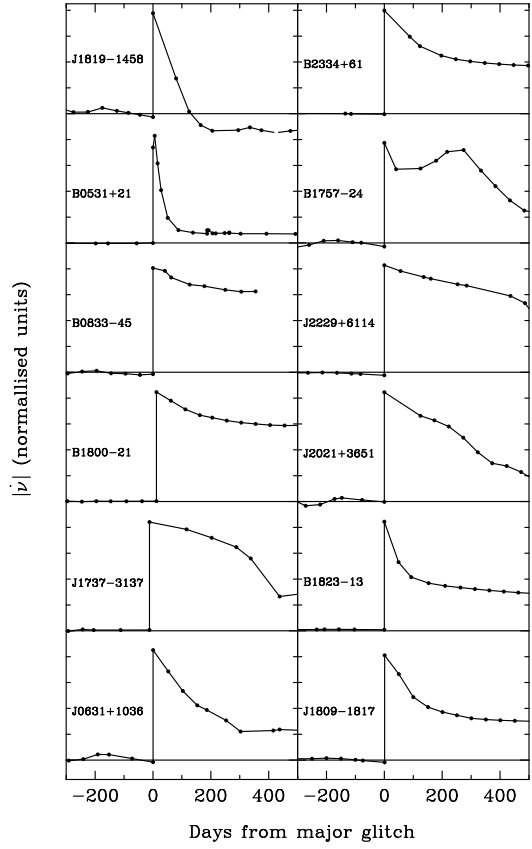} 
  \end{center}   
  \vspace{-15pt}
  \caption{\small{The relative change in the magnitude of $\dot{\nu}$
  due to glitches, in J1819$-$1458 and a sample of glitches in other
  pulsars \cite{lmk+09}.}}
%  \vspace{-15pt}
  \label{fig:f1}
\end{figure}

The original RRAT sources have now been monitored for several
years. This has led to coherent timing solutions and in the case of
J1819$-$1458 the detection of glitches. Glitches are step changes in
spin frequency $\nu$ and its derivative $\dot{\nu}$ of the form
\begin{equation}
  \nu(t)\rightarrow\nu(t)+\Delta\nu_{\mathrm{p}}+\Delta\dot{\nu}_{\mathrm{p}}t+\Delta\nu_{\mathrm{d}}e^{-t/\tau_{\mathrm{d}}}
\end{equation}
\begin{equation}
  \dot{\nu}(t)\rightarrow\dot{\nu}(t)+\Delta\dot{\nu}_{\mathrm{p}}+\Delta\dot{\nu}_{\mathrm{d}}e^{-t/\tau_{\mathrm{d}}}
\end{equation}
where the permanent steps are labelled with a `p' and the steps
labelled `d' decay on a timescale of $\tau_{\mathrm{d}}$
\cite{sl96}. The glitches detected in J1819$-$1458 have fractional
sizes of $\Delta\nu/\nu=0.6\times10^{-6}$ and $0.1\times10^{-6}$,
similar in size to those seen in young pulsars \cite{lmk+09}. The
noteworthy point however is that the net effect of the glitches is to
decrease the slow-down rate of the star's rotation, i.e. the magnitude
of $\dot{\nu}$ decreased. This is completely anomalous and unlike all
radio pulsars glitches ever detected (see Figure~\ref{fig:f1}). When
contemplating the significance of this effect we might consider that
the effect of the glitches in $P-\dot{P}$ space is to move
J1819$-$1458 (labelled in Figure~\ref{fig:ppdot}) downwards. If we
were to propose that such glitches were typical in this source then it
would suggest that J1819$-$1458 previously occupied the region of
$P-\dot{P}$ space where the magnetars are. The importance of such
effects must be considered when we consider pulsar (and magnetar) spin
evolution in the $P-\dot{P}$ diagram.

\section{RRATs: Special or Not?}
There has been some debate about what exactly a RRAT is and if in fact
they are `special' or not. Here we investigate these
questions. Firstly we invoke the (effective) definition implemented
for RRATs discovered in the PMPS \cite{mll+06,kle+10}: a RRAT is a
source identified in a single pulse (SP) search rather than a
periodicity (FFT) search. If we let $r$ denote the ratio of the SP and
FFT search signal-to-noise ratios,
i.e. $r=(S/N)_{\mathrm{SP}}/(S/N)_{\mathrm{FFT}}$, then RRATs are
sources with $r>1$. We see immediately that this definition depends on
observing time as well as being at the mercy of the (a priori unknown)
pulse amplitude distributions of RRATs. The definition is a detection
classification only, i.e. sources identified as ``RRATs'' ($r>1$) in
one survey may well be identified as ``pulsars'' ($r<1$) in another
survey with longer pointings. Are the group of RRATs, so defined, in
any way special?

\begin{figure}                                                          
  \begin{center} 
    \includegraphics[scale=0.4,angle=-90]{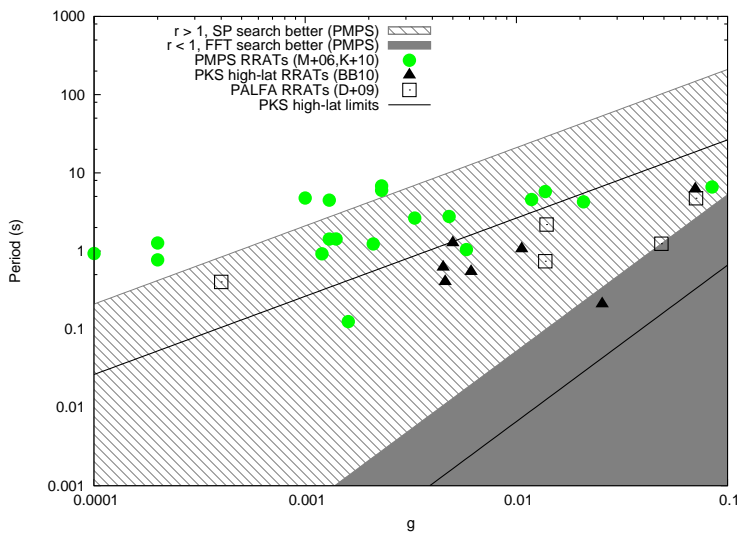}       
  \end{center}   
  \vspace{-15pt}                                                     
  \caption{\small{Plotted here is $g-P$ space with the regions where
  SP searches (hatched) and FFT searches (shaded) are more effective
  for the PMPS \cite{mlc+01}, which define the ``RRAT'' and ``pulsar''
  regions. Over-plotted are the PMPS RRATs with measured periods as
  reported in \cite{mll+06,kle+10} (M+06 and K+10 in the figure). Also
  plotted are the ``RRAT'' and ``pulsar'' boundaries for the Parkes
  high-latitude surveys and the sources discovered therein which have
  known $P$ and $g$ \cite{bb10} (BB10 in the figure). For J1654$-$23
  we use the correct period as recently determined, not that published
  in BB10 \cite{keane_inprep}. We also plot the sources reported in
  \cite{dcm+09} (D+09 in the figure). J1854+03 is plotted with the
  PMPS sources, although it was also identified in PALFA. We note that
  the boundaries for the inner-Galaxy PALFA pointings are the same as
  for the Parkes high-latitude surveys if we assume no difference in
  sensitivity. This is of course incorrect, and due to this extra
  difference (the Parkes surveys have the same sensitivity as each
  other) the D+09 sources are plotted simply for illustration.}}
  \label{fig:gP}
\end{figure}

To answer this we consider what this definition means as far as
selection effects are concerned. If we take a source which emits
pulses a fraction $g$ of the time and nulls a fraction $1-g$ of the
time then we can derive the condition for $r>1$ to be
$N^{-1}<g<2N^{-1/2}$ where $N$ is the number of pulse periods during
our observation and we have ignored some pulse shape factors of order
unity \cite{mc03}. If we observe for a time $T=NP$ then we can convert
this to a constraint on $g-P$ space which is $Tg^2/4<P<Tg$
\cite{mlk+09}. For a given $g$, the low period limit defines the $r=1$
condition so that at lower periods an FFT search is more
effective. For higher periods than $Tg$ there is unlikely to be even
one pulse during the observation. Thus RRATs are those sources
detected in the hatched region in $g-P$ space in
Figure~\ref{fig:gP}. Clearly this definition depends on the observing
time $T$, and the boundaries shown in Figure~\ref{fig:gP} are for the
35-minute pointings of the PMPS \cite{mlc+01}. Different surveys will
have different `RRAT-PSR' boundaries, e.g. the higher-latitude Parkes
surveys \cite{bb10} had shorter pointings and hence different
boundaries which are over-plotted on Figure~\ref{fig:gP}. Thus the
``RRAT'' J1647$-$36 detected in the high-latitude surveys would have
been detected as a ``pulsar'' if it were surveyed in the PMPS. We note
that in reality the $g$ values we measure represent the
\textit{apparent} nulling fraction, i.e. the intrinsic values of $g$
may be higher depending on the pulse-to-pulse modulation and distance
to the source \cite{wsrw06,bb10}.

So the definition of a RRAT is arbitrary, survey-dependent and makes a
selection in $g-P$ space. FFT searches also make a selection in $g-P$
space but, in comparison, single pulse searches are sensitive to
higher period sources (up to several seconds) with moderate nulling
fraction down to very short period pulsars with large nulling
fractions. It seems unfair to compare period distributions of sources
selected in this way, but we do note that the periods of many PMPS
RRATs are well above (more than an order of magnitude) the minimum
periods where they would still be classified as RRATs, i.e. for a
given $g$ the PMPS was sensitive to low-period RRATs
(e.g. $P\lesssim1.0$~s and $g\lesssim0.001$) but these were not
detected. This is also true for most of the high-latitude survey
sources and boundaries. Monitoring RRATs over some time reveals their
slow-down rate $\dot{P}$ which is not subject to any selection
effect. This has shown that RRATs have high spin-down rates compared
to the normal radio pulsars \cite{mll+06,mlk+09,keane_inprep} and this
implies stronger magnetic fields according to
Equation~\ref{eq:B}. This suggests the question of whether long period
and/or high $B$ sources have higher nulling fractions or modulation
indices (i.e. low observed $g$ values). Here we reach a dead end
because, as discussed in \S~2.2, the nulling properties of pulsars,
i.e. the $g-P$ distribution of FFT-selected sources, is unknown. A
project to determine the nulling characteristics of a large population
of pulsars may shed some light on how $g$ depends on pulsar parameters
like $P$ and $B$. A weak correlation of modulation index with $B$ has
been suggested in \cite{wes06}.

As discussed in \cite{bb10}, there are three credible explanations for
the nature of RRATs: (1) a distinct population with high nulling
fraction; (2) a distinct evolutionary phase with high nulling fraction
(dubbed ``true RRATs''); (3) weak/distant pulsars with a high
modulation index. There seems to be no reason to consider RRATs as a
distinct population. In fact, as discussed in \S~3.1, this leads to
inconsistencies \cite{kk08}. Solutions (2) and (3) are both consistent
with high observed nulling fractions, i.e. low values of $g$. The
question of the ``RRAT emission mechanism'', for which there have been
many proposed explanations \cite{li06,zgd07,lm07,cs08,olny09}, might
then more accurately be re-phrased as a question of what causes
nulling of the pulsar emission mechanism and how this might occur on
long ($\sim10-10^4$ period) timescales. The high projected population
of RRATs also becomes less if some sources are covered by solution
(3). Such sources will have low-luminosity\footnote{Here 'luminosity'
is used to refer to the quantity $L=SD^2$ which has units of
$\rm{Jy.kpc}^2$ or alternatively $W.Hz^{-1}$.} periodic emission. The
pulsar population is estimated only above some threshold luminosity
$L_{\mathrm{min}}\sim 0.1\;\rm{Jy.kpc}^2$ (for periodic emission), so
that if these sources are above $L_{\mathrm{min}}$ they are already
accounted for within low-luminosity selection effect scaling factors
in estimates of the pulsar population (see e.g. \cite{lfl+06}). If the
underlying periodic emission were below $L_{\mathrm{min}}$ then these
sources would contribute to a birthrate problem by increasing the
pulsar population estimate, and indeed the required low-luminosity
turn-over\footnote{There must be a low-luminosity turn-over so that
the integral $\int N(L)dL$ does not diverge at the low end. Here
$N=N(L)$ denotes the number of pulsars with luminosity between $L$ and
$L+dL$.} is not yet seen, which is why artificial cut-offs are usually
applied in population syntheses (see e.g. \cite{fk06}). Extreme
modulation can account for all but two RRATs, according to the
analysis of \cite{bb10} (but notably not J1819$-$1458, which agrees
with \cite{josh_inprep}), but the true number may be larger as it
assumes analogues of the extreme source PSR~B0656+14 to be common in
the Galaxy. RRAT pulse amplitude distributions will shed more light on
these matters \cite{josh_inprep}. Another recently discovered
phenomenon is sources switching between RRAT-like and pulsar-like
modes like PSR~J0941$-$39 \cite{bb10}, consistent with the suggestion
that nulling is a type of moding \cite{wmj07}, although it is unknown
how this correlates with pulsar parameters such as age, $P$ or
$B$. This also leads to the suggestion that nulling fraction may
increase in steps rather than gradually as a pulsar evolves.

\section{Conclusion}
The transient radio sky at millisecond scales contains thousands of
neutron stars. Searches for isolated bursts (rather than FFT searches
for periodic emission) have revealed sources which appear to null on
various timescales. Much recent interest has focused on the
RRATs. These sporadically emitting sources are abundant in the Galaxy
so that it seems necessary to absorb the RRATs within the known
neutron star populations \cite{kk08}, either as a distinct
evolutionary stage or as pulsars with extreme pulse amplitude
variability. As the RRATs have longer periods than might be expected
(given the selection effects) and higher $B$ values than the normal
radio pulsars, this suggests that long $P$ and/or high-$B$ sources may
have increased nulling fractions and/or increased pulse-to-pulse
modulation, in comparison to the general (FFT-selected) pulsar
population. Of the RRATs which do not seem to fit the mould of highly
modulated pulsars, J1819$-$1458 is the best studied source (and the
source with the strongest $B$). It has been seen to undergo anomalous
glitches. These have been proposed to have had an evolutionary impact
so that J1819$-$1458 may be an exhausted magnetar \cite{lmk+09},
although the importance of this unique glitch behaviour is still
uncertain. Recently sources have been identified which alternate
between RRAT-like and pulsar-like behaviour, perhaps transitionary
objects. The class of intermittent pulsars has lengthy nulls of
several days which are quasi-periodic. It is difficult to understand
these periodicities which are apparently absent from the RRATs
\cite{nipuni_inprep}. More information on nulling phenomenology, such
as the `periodic nulls' in PSR~J1920+1040 and the `component nulls' in
PSR~J1326$-$6700 reported by \cite{wmj07}, is needed to relate these
transient behaviours to one another.

These discoveries also highlight our lack of knowledge of neutron star
evolution post-supernova. The XDINSs and magnetars must also be
accommodated into any evolutionary paradigm. It is important to point
out that the beaming fraction of long period sources is small. If we
extrapolate the empirical pulsar beaming fraction \cite{tm98}, which
was derived from measurements of low-period ($P<2\;$s) pulsars, to
RRATs and XDINSs (periods up to 11~s), we get very small values of
$f_{\mathrm{beam}}\approx0.03$. Thus the chance of missing any beamed
radio emission from the 7 well studied XDINS is high at
$\sim(0.97)^7\approx0.8$ and we should not dismiss large beaming
effects for these long-period sources. We also wish to explain the
origin of magnetars. The standard pulsar spin-down model presented in
\S~2 with dipole braking, i.e. $n=3$, is not realised in the handful
of sources with known braking indices. In fact the measured values of
$n$ range from $-1.5$ to $2.9$
\cite{lps93,lpgc96,ckl+00,mmw+06,lkg+07}, telling us that
$\tau_{\mathrm{c}}$ is an unreliable age estimate. Unfortunately
alternative age estimates, such as cooling ages, kinematic ages or
supernova remnant associations, are known only for a small number of
sources. The low braking indices also imply increasing surface
magnetic fields strengths (clear from combining the spin-down law with
Equation~\ref{eq:B}) and there is also some evidence for magnetic
field alignment with the rotation axis \cite{wj08}. While these
effects are poorly understood they must be considered (see
e.g. \cite{rl10}) to determine a full picture of neutron star
evolution.

Further progress in pursuit of these questions will be made with the
identification of many more sources. New radio transients, like those
described here, are expected to be detected in abundance with next
generation instruments like LOFAR, the ATA, FAST, the SKA pathfinders,
and, in a decade or so, with the SKA itself (see \cite{lbb+09} and
references therein).

\section*{Acknowledgements}
EK thanks M.~B. Purver for helpful comments and acknowledges the
support of a Marie-Curie EST Fellowship with the FP6 Network
``ESTRELA'' under contract number MEST-CT-2005-19669.

% BIBTEX STUFF
%\bibliographystyle{plain_copy}
%\bibliography{journals,journals_apj,psrrefs,modrefs,crossrefs}

% PASTED IN BBL FILE

\end{document}